\documentclass{article}
\usepackage{spconf,amsmath,graphicx,hyperref}
\usepackage{cite}
\usepackage{amsmath,amssymb,amsfonts}
\usepackage{graphicx}
\usepackage{textcomp}
\usepackage{xcolor}
\usepackage{amsmath,graphicx,booktabs,multirow,multicol,color}
\usepackage{algorithmicx, algpseudocode}
\usepackage[linesnumbered,ruled,vlined]{algorithm2e}

\usepackage[labelfont=bf]{caption} 
\renewenvironment{thebibliography}[1]{
  \begin{oldthebibliography}{#1}
  \setlength{\itemsep}{0.em}
  \setlength{\parskip}{0.em}
}
{
  \end{oldthebibliography}
}

\title{MeanSE: Efficient Generative Speech Enhancement with Mean Flows}
\name{
\begin{tabular}{c}
Jiahe Wang$^{1}$, Hongyu Wang$^{1}$, Wei Wang$^{1}$, Lei Yang$^3$, Chenda Li$^{1}$, \\
Wangyou Zhang$^{2}$, Lufen Tan$^3$, Yanmin Qian$^{1}$ \\
\end{tabular}
}
\address{$^1$Auditory Cognition and Computational Acoustics Lab \\
MoE Key Lab of Artificial Intelligence, AI Institute \\
School of Computer Science, Shanghai Jiao Tong University, Shanghai, China \\
$^2$School of Artificial Intelligence, Shanghai Jiao Tong University, Shanghai, China \\
$^3$Samsung R\&D Institute China - Beijing (SRC-B)}

\begin{document}
%
\maketitle

\begin{abstract}
Speech enhancement (SE) improves degraded speech's quality, with generative models like flow matching gaining attention for their outstanding perceptual quality. However, the flow-based model requires multiple numbers of function evaluations (NFEs) to achieve stable and satisfactory performance, leading to high computational load and poor 1-NFE performance. In this paper, we propose MeanSE, an efficient generative speech enhancement model using mean flows, which models the average velocity field to achieve high-quality 1-NFE enhancement. Experimental results demonstrate that our proposed MeanSE significantly outperforms the flow matching baseline with a single NFE, exhibiting extremely better out-of-domain generalization capabilities.
\end{abstract}


\section{Introduction}
\label{sec:intro}

Speech Enhancement (SE) \cite{SEreview,SEsurvey} is a fundamental audio processing task focused on improving the quality and intelligibility of speech signals degraded by noise, reverberation, or interference. Generative models \cite{sgmse,conddiffse,storm,causaldiffSE,flowse,flowse_text,modifyingflowmatching,twoflow} have gained increasing attention in the domain of speech enhancement. Compared to traditional discriminative SE models \cite{SE1,SE2}, generative models not only yield superior perceptual quality, but also demonstrate stronger generalization capabilities \cite{conddiffse}.

Among them, score-based generative models \cite{sgmse,conddiffse,storm,causaldiffSE} first demonstrated notable success by estimating clean speech through solving the reverse stochastic differential equations (SDE) \cite{scorematching}.
However, the number of function evaluations (NFEs), which represents how many times the neural network is run in the inference process, is usually high (\textit{e.g.}, 30 NFEs in \cite{sgmse} and 50 NFEs in \cite{conddiffse}), thus brings about huge computational load.
Subsequently, FlowSE \cite{flowse,flowse_text,modifyingflowmatching,twoflow}, based on conditional flow matching (CFM) \cite{flowmatching}, emerged as a lightweight alternative. FlowSE learns a continuous, deterministic transformation (the `flow') defined by the time-dependent velocity field, mapping a simple probability distribution (\textit{e.g}., noisy speech) to a complex target data distribution (\textit{e.g.}, clean speech) conditionally,
thus achieves better performance with fewer NFEs by solving the ordinary differential equations (ODE) \cite{chen2018neural}.
However, FlowSE still requires several NFEs (\textit{e.g.} 5 NFEs in \cite{flowse} and 10-20 NFEs in \cite{flowse_text}) to achieve stable performance, while the 1-NFE performance often degrades significantly. This makes the flow matching-based approach still computationally expensive and unsuitable for real-world applications with high computational restrictions, such as real-time speech enhancement \cite{realtimeSE1,realtimeSE2}. Though recent work \cite{modifyingflowmatching} has achieved considerable 1-NFE enhancement performance on intrusive metrics and WER through the idea of data prediction \cite{data_prediction}, this discriminative-relied method may lose the characteristics of generative models, leading to worse performance on non-intrusive metrics.

In this paper, we propose MeanSE\footnote{Source code on: \url{https://github.com/Twinkzzzzz/MeanSE}}, an efficient generative speech enhancement model with mean flows \cite{meanflow,meanaudio}. The mean flow models the average velocity field rather than the instantaneous velocity field in flow matching. This distinction is crucial as it enables high-quality speech enhancement with only 1 NFE by predicting the average velocity of the whole transformation in the flow.
Experimental results demonstrate that, compared to the flow matching-based baseline, MeanSE significantly outperforms the baseline with 1 NFE on the in-domain test, and comprehensively exhibits superior generalization capabilities across different NFEs on the out-of-domain test. To the best of our knowledge, this paper is the first attempt to apply mean flows to speech enhancement.

\section{Method}
\label{sec:method}

\subsection{Speech Enhancement with Flow Matching}
\label{section:flow}

We take FlowSE \cite{flowse} as the baseline in this paper. FlowSE adopts the conditional flow matching (CFM) \cite{flowmatching} for speech enhancement. Given the noisy speech $y$, the ODE incorporates $y$ as the condition to form a flow:
\begin{equation}
\label{eq:flow}
    \frac{dx_t}{dt} = v_t(x_t|y),~x_t = \phi_t(x_0|y),
\end{equation}
where $t \in [0,1]$ denotes the time step, $\phi_t:[0,1] \times \mathbb{R}^d \to \mathbb{R}^d$ denotes the flow defined by the time-dependent instantaneous velocity field $v_t:[0,1] \times \mathbb{R}^d \to \mathbb{R}^d$, $d$ denotes the data dimension. The flow generates the probability path $p_t(x_t|y)$ from $p_1(x_1|y) = \mathcal{N}( y, \sigma^2 \mathbf{I})$ to $p_0(x_0|y)$, where the former is the distribution of noisy speeches and the latter is trained to approximate the distribution of clean speeches. $\sigma$ is a hyperparameter denoting the variance in the Gaussian distribution of $p_1(x_1|y)$.
The training objective is to estimate the velocity field $v_t(x_t|y)$ by a neural network $v_{\theta} (x_t,y,t)$ using the CFM loss. The clean speech $x_0$ can be generated by solving the ODE in \ref{eq:flow}. Given a pair of training samples $(x_0,y)$, the CFM loss can be formulated as follows:
\begin{equation}
    \mathcal{L}_{CFM} = \mathbf{E}_{t, (x_0,y), p_1(x_1|y)} \lVert v_{\theta} (x_t,y,t) - v_t(x_t|x_0,y)  \rVert^2,
    \label{eq:cfmenhancement}
\end{equation}
Following \cite{flowmatching,flowse}, we set $p_t(x_t|x_0,y)$ as the Gaussian conditional probability path. The conditional flow $x_t$ and the conditional velocity field $v_t(x_t|x_0,y)$ can be formulated as:
\begin{equation}
    x_t = \frac{\sigma_t }{\sigma_1 }\left(x_1-\mu_1(x_0,y)\right)+\mu_t(x_0,y),
    \label{eq:sample_x_t}
\end{equation}
\begin{equation}
    v_t(x_t|x_0,y) = \frac{\sigma_t^\prime }{\sigma_t }(x_t-\mu_t(x_0,y))+\mu_t^\prime(x_0,y),
    \label{eq:sample_v_t}
\end{equation}
where $x_1$ is sampled from $p_1(x_1|x_0,y)$. The mean $\mu_t(x_0,y)$ and variance $\sigma_t^2$ for $p_t(x_t|x_0,y)= \mathcal{N}\left( \mu_t(x_0, y), \sigma_t^2 \mathbf{I}\right)$ are configured as follows: 
\begin{equation}
    \mu_t(x_0,y) = (1-t)x_0+ty,
    \label{eq:mean} 
\end{equation}
\begin{equation}
    \sigma_t=t\sigma,
    \label{eq:std}
\end{equation}
In the training process, given a training pair $(x_0, y)$, we first sample $t$ uniformly from $\mathcal{U}(0,1]$. Then we sample $x_t$ from the aforementioned $p_t(x_t|x_0,y)$. The training objective is constructed as Equation \eqref{eq:cfmenhancement}.
In inference, we randomly sample $x_1$ from $p_1(x_1|x_0,y)$, and then solve the ODE defined below:
\begin{equation}
    dx_t = v_{\theta} (x_t, y, t)dt,
\end{equation}
\begin{equation}
    \label{eq:CFM_int}
    x_0 = x_1 + \int^0_1v_{\theta}(x_t, t, y)dt,
\end{equation}
where numerical solvers like Euler Method \cite{flowse,meanaudio} are implemented to approximate this integration.

\subsection{Speech Enhancement with Mean Flows}
\label{section:meanflow}
We propose MeanSE, which applies mean flows \cite{meanflow, meanaudio} to speech enhancement. Unlike standard flow matching that models the instantaneous velocity field, the mean flow models the average velocity field between two time steps in the latent space.
A time interval $[r, t], 0 \leq r < t \leq 1$ is given, where the average velocity $u(x_t, r, t|y)$ within the interval $[r, t]$ conditioned on the noisy speech $y$ can be easily calculated by the integration of the instantaneous velocity field:
\begin{equation}
    u(x_t, r, t|y) = \frac{1}{t-r} \int_{r}^{t} v_{\tau}(x_{\tau}|y) \, d\tau,
\end{equation}
where $x_{\tau}$ represents the state in the flow at time step $\tau$. By differentiating both sides of the equation with respect to $t$, and through operations such as Leibniz's rule for integration, we can derive the following formulation:
\begin{equation}
    u(x_t, r, t|y) = v_t(x_t|y) - (t - r) \frac{d}{dt}u(x_t, r, t|y),
\end{equation}
where the flow $x_t$ and the instantaneous velocity $v_t$ can be sampled in exactly the same way as that in Equation \ref{eq:sample_x_t} and Equation \ref{eq:sample_v_t}.
The training objective is to model the average velocity field $u(x_t, r, t|y)$ by a neural network $u_{\theta}(x_t, r, t, y)$, the loss function is thus given by:
\begin{equation}
    \mathcal{L}_{MF} = \mathbf{E}_{t, (x_0,y), p_1(x_1|y)} \|u_{\theta}(x_t, r, t, y) - \text{sg}(u(x_t, r, t|y))\|^2,
\end{equation}
where $\text{sg}(\cdot)$ denotes the stop-gradient operation.

In the inference stage, supposing the NFE is set to $N$, the time integral in Equation \ref{eq:CFM_int} can be replaced by dividing the time interval $[0,1]$ into $N$ sub-intervals $\{[r_i, t_i]|i \in \{0,...k-1\}\}$ evenly and predicting the average velocity in each sub-interval. The whole process can be presented as the following pseudo-code:

\begin{algorithm}[htbp]
\SetAlgoLined
\caption{MeanSE Inference}
\label{alg:inference}
\KwIn{Trained network $u_{\theta}$, noisy speech $y$, NFE=$N$}
\KwOut{The enhanced speech $x_0$}
Sample $x_1 \sim \mathcal{N}\left( \mu_1(x_0, y), \sigma_1^2 \mathbf{I}\right)$ \\
$x_t = x_1$ \\
\For{$i = N-1$ to $0$}{
    $r_i = \frac{i}{N},~t_i = \frac{i+1}{N}$ \\
    $x_t = x_t + (r_{i} - t_{i}) \cdot u_{\theta}(x_t, r_i, t_i, y)$
}
$x_0 = x_t$
\end{algorithm}
In particular, the 1-NFE inference can be simplified to:
\begin{equation}
    x_0 = x_1 - u_{\theta}(x_1, 0, 1, y).
\end{equation}

\subsubsection{Fusing Time Steps in the Neural Network}

Different from FlowSE, the neural network in MeanSE accepts two time steps $t$ and $r$ as input. These two time steps pass through the same Gaussian Fourier transform and linear layers, obtaining two $K$-dimensional time embeddings. Subsequently, these two time embeddings will be concatenated to form a $2K$-dimensional feature. An additional linear layer is applied to map it back to a $K$-dimensional fused feature. This $K$-dimensional feature is then fused with the speech spectrum in the same way as that in FlowSE.

\subsubsection{Stabilizing the Flow Fields}
\label{sec:stabilize}

In the experiment, We found that directly forcing the model to regress the average velocity field from scratch makes the training process unstable and hard to converge. MeanSE addresses this issue by leveraging the following strategies:

\textbf{Flow Field Mix-up}: Following prior works \cite{meanflow, meanaudio}, we adopt the flow field mix-up strategy. During the training stage, the model is set to sample $t=r$ at a certain `flow ratio', which force the neural network to include the learning of the instantaneous velocity field as in flow matching. We include the ablation study of this flow ratio in Section \ref{sec:flow_ratio}.

\textbf{Time Interval Curriculum Learning}: During training, we found that the model struggles to directly learn about the average velocity in large time intervals $[r,t]$, and potentially fails to effectively estimate the average velocity. To address this issue, we apply a multi-stage time interval curriculum learning approach. The model is initialized by the trained FlowSE checkpoint. During training, we gradually increase the maximum sampling interval for $[t,r]$, from 0.2, through 0.4, 0.6, 0.8, finally to 1. In each stage, the model is fine-tuned based on the initialization by the trained checkpoint in the previous stage. This enables the model to progressively adapt to learning from small to large intervals, providing a stable training process.

\section{Experiment}
\label{sec:exp}
\subsection{Dataset}
\label{sec:dataset}

We choose the VoiceBank-DEMAND \cite{VoiceBank,DEMAND} dataset and the WHAMR! \cite{WHAMR} dataset to test the in-domain performance and out-of-domain generalization capabilities of our proposed model, respectively.
The VoiceBank-DEMAND dataset is constructed by combining clean speech samples from the VCTK Corpus \cite{VoiceBank} with interfering sounds from the DEMAND \cite{DEMAND} dataset. For the training subset, mixtures are generated at signal-to-noise ratios (SNRs) of 0, 5, 10, and 15 dB, while the test set uses SNRs of 2.5, 7.5, 12.5, and 17.5 dB. Following previous works \cite{sgmse,storm}, we divide the dataset into a 26-speaker training set (10802 utterances), a 2-speaker validation set (770 utterances, speaker `p226' and `p287'), and a 2-speaker test set (824 utterances).

WHAMR! \cite{WHAMR} is an extension of the WHAM! \cite{WHAM} dataset that incorporates synthetic reverberation to simulate realistic indoor acoustic environments. We utilized the \textit{mix\_single\_anechoic} version of test set, which combines clean speeches from the WSJ0 Corpus \cite{WSJ0} with real-recorded ambient noise and room impulse responses.
Both the FlowSE baseline and our propose MeanSE are trained on the VoiceBank-DEMAND dataset, and tested on both VoiceBank-DEMAND and WHAMR!.
The sampling rate is set to 16 kHz for all the speech signals used in the experiment.

\subsection{Model Setups and Training Details}
\label{sec:training}

Following prior works \cite{sgmse,storm,conddiffse,flowse}, we adopt the Noise Conditional Score Network (NCSN++) in both the FlowSE and the MeanSE.
The speech signals are processed in the form of complex spectrum of Short-time Fourier transform (STFT), where $n_{fft}$ and \textit{hop length} are set to 1022 and 320. The \textit{image size} of the NCSN++ is set to 512.

When initializing the MeanSE model with the trained FlowSE checkpoint,
we manually initialize the linear layer fusing $t$ and $r$, enabling it to work as the trained FlowSE model. Its weight matrix is initialized by the concatenation of a $K \times K$ zero matrix and a $K \times K$ diagonal matrix, and its bias is initialized by a $K$-dimensional zero vector, where $K$ denotes the dimension of the time embeddings. This makes the fused time embedding equal to the original embedding of $t$. Then the MeanSE model is fine-tuned applying the aforementioned time interval curriculum learning.

In terms of other training configurations, the learning rate is set to $1 \times 10^{-4}$ initially in from-scratch experiments and $1 \times 10^{-5}$ in fine-tune experiments, with the weight decay set to $1 \times 10^{-6}$. The $\sigma$ in the ODE is set to 0.5. The batch size is set to 2.
Each model is trained for about 40 epoches until converge, and the checkpoint with the lowest validation loss is chosen for testing.

\subsection{Evaluation Metrics}
\label{sec:evaluation}

We employ several metrics to comprehensively evaluate the performance of the proposed FlowSE. Two intrusive metrics are utilized, where PESQ \cite{PESQ} predicts overall speech quality by comparing the degraded signal to a clean reference using a psychoacoustic model, and ESTOI \cite{STOI} estimates the speech intelligibility by analyzing the correlation between the temporal envelopes of the clean and processed signals in the time-frequency domain. Four additional non-intrusive metrics: DNSMOS \cite{DNSMOS} (DNSMOS P.835 including three measurements of SIG, BAK and OVRL), WVMOS \cite{WVMOS}, UTMOS \cite{UTMOS} and NISQA \cite{NISQA}, are utilized to comprehensively measure the human-perceived naturalness, overall quality and credibility of the enhanced speech signals.

\subsection{Experimental Results}
\label{sec:results}

\subsubsection{In-domain performance}
\label{sec:in_domain}

\begin{table*}[t!]
\centering
\setlength{\tabcolsep}{10.6pt}
\normalsize
\label{tab:in_domain}
\caption{The in-domain performance comparison between our proposed MeanSE and the baseline FlowSE. Both models are trained and tested on the VoiceBank-DEMAND \cite{VoiceBank,DEMAND} dataset. The flow ratio is set to 0.75 in MeanSE. \iffalse The numbers in bold indicate better performances comparing two models under the setting of 1-NFE.\fi}
\label{tab:in_domain}
\begin{tabular*}{\textwidth}{c|c|ccc|c|c|c|c|c}
\toprule
\textbf{Model} & \textbf{NFE} & \textbf{SIG} & \textbf{BAK} & \textbf{OVRL} & \textbf{UTMOS} & \textbf{WVMOS} & \textbf{NISQA} & \textbf{PESQ} & \textbf{ESTOI} \\
\midrule
\multirow{3}{*}{FlowSE} & 5 & 3.327 & 3.879 & 2.992 & 3.596 & 3.954 & 3.402 & 2.347 & 0.804 \\
 & 2 & 3.294 & 2.868 & 2.961 & 3.585 & 3.913 & 3.394 & 2.366 & 0.791 \\
 & 1 & 3.336 & 3.177 & 2.685 & 3.317 & 3.375 & 3.420 & 1.843 & 0.761 \\
\midrule
\multirow{3}{*}{MeanSE} & 5 & 3.332 & 3.874 & 2.997 & 3.567 & 3.898 & 3.282 & 2.347 & 0.819 \\
 & 2 & 3.320 & 3.850 & 2.976 & 3.578 & 3.859 & 3.348 & 2.297 & 0.821 \\
 & 1 & 3.317 & 3.528 & 2.841 & 3.483 & 3.644 & 3.552 & 2.090 & 0.800 \\
\bottomrule
\end{tabular*}
\end{table*}

Table \ref{tab:in_domain} compares the in-domain performance of our proposed MeanSE against the baseline FlowSE on the VoiceBank-DEMAND dataset, evaluating both models across different NFEs. The flow ratio is set to 0.75 in MeanSE.
The experimental results show that, on the multi-NFE (\textit{i.e.}, NFE=5 and 2) conditions, MeanSE demonstrates comparable performance to FlowSE, with MeanSE performing better on SIG, OVRL and ESTOI and FlowSE scoring higher on BAK, UTMOS, WVMOS and NISQA.
However, when NFE is set to 1, MeanSE substantially outperforms FlowSE across nearly all the metrics, where SIG is the only exception. This proves that MeanSE performs much better on the scenario of extremely low computational load, showing greater potential to applications such as real-time speech enhancement \cite{realtimeSE1,realtimeSE2}.

\subsubsection{Generalization to out-of-domain dataset}

Table \ref{tab:out_domain} evaluates the generalization capabilities of MeanSE and FlowSE by testing both the models on WHAMR! \cite{WHAMR} dataset after training on VoiceBank-DEMAND \cite{VoiceBank,DEMAND}. The header `DNSMOS' in the table represents the OVRL criterion. The flow ratio is set to 0.75 in MeanSE. The results presented in the table show that, across all the investigated NFEs, MeanSE demonstrates a stable and significant performance improvement. The most striking advantage emerges when the NFE is set to 1, where FlowSE seems to virtually fail to generate valid enhanced speech signals, however MeanSE works effectively and exhibits dramatic improvements over FlowSE (+0.363 on DNSMOS, +0.402 on UTMOS, +1.090 on WVMOS and +0.511 on NISQA).
Such results underscore the excellent generalization capability of MeanSE, especially under low-computation constraints.

\begin{table}[t]
\centering
\normalsize
\setlength{\tabcolsep}{3pt}
\caption{The comparison of generalization capabilities of MeanSE and FlowSE. The models are trained on VoiceBank-DEMAND \cite{VoiceBank,DEMAND} and tested on WHAMR! \cite{WHAMR}. The flow ratio is set to 0.75 in MeanSE.}
\label{tab:out_domain}
\begin{tabular*}{\linewidth}{c|c|cccc}
\toprule
\textbf{Model} & \textbf{NFE} & \textbf{DNSMOS} & \textbf{UTMOS} & \textbf{WVMOS} & \textbf{NISQA} \\
\midrule
\multirow{3}{*}{FlowSE} & 5 & 2.331 & 1.742 & 1.948 & 2.102 \\
 & 2 & 2.262 & 1.720 & 1.582 & 2.043 \\
 & 1 & 1.785 & 1.522 & 0.922 & 2.012 \\
 \midrule
\multirow{3}{*}{MeanSE} & 5 & 2.441 & 1.966 & 2.220 & 2.494 \\
 & 2 & 2.417 & 2.013 & 2.257 & 2.617 \\
 & 1 & 2.148 & 1.924 & 2.012 & 2.523 \\
\bottomrule
\end{tabular*}
\end{table}

\subsubsection{Ablation Study of Flow Field Mix-up}
\label{sec:flow_ratio}

Finally, we investigate the effectiveness of the flow field mix-up with different flow ratios.
Four different flow ratios: 0.0, 0.25, 0.5 and 0.75 are investigated in the ablation study. For simplicity, the results with 1 NFE are taken as examples. As presented in Table \ref{tab:flow_ratio}, on both the VoiceBank-DEMAND and WHAMR! datasets, increasing the flow ratio from 0.0 to 0.75 consistently brings about performance improvements, where the only exception is DNSMOS on the VoiceBank-DEMAND dataset. Moreover, setting the flow ratio to 0.0 hardly yields valid results, as the training process becomes highly unstable and hard to converge. These experimental results coincide with the findings in prior work \cite{meanaudio}.

\begin{table}[t]
\centering
\normalsize
\setlength{\tabcolsep}{2pt}
\caption{Ablation study on the flow ratio (NFE is set to 1).}
\label{tab:flow_ratio}
\begin{tabular*}{\linewidth}{c|c|cccc}
\toprule
\textbf{Dataset} & \textbf{Ratio} & \textbf{DNSMOS} & \textbf{UTMOS} & \textbf{WVMOS} & \textbf{NISQA} \\
\midrule
\multirow{4}{*}{VCTK} & 0.0 & 2.678 & 3.295 & 3.354 & 3.424 \\
 & 0.25 & \textbf{2.843} & 3.465 & 3.624 & 3.525 \\
 & 0.5 & 2.826 & 3.466 & 3.615 & 3.550 \\
 & 0.75 & 2.841 & \textbf{3.483} & \textbf{3.644} & \textbf{3.552} \\
\midrule
\multirow{4}{*}{WHAMR!} & 0.0 & 1.775 & 1.507 & 0.791 & 1.969 \\
 & 0.25 & 2.114 & 1.895 & 1.968 & 2.486 \\
 & 0.5 & 2.140 & 1.900 & 1.976 & 2.520 \\
 & 0.75 & \textbf{2.148} & \textbf{1.924} & \textbf{2.012} & \textbf{2.523} \\
\bottomrule
\end{tabular*}
\end{table}

\section{Conclusion}
\label{sec:conclusion}

In this paper, we propose MeanSE, a generative speech enhancement (SE) model based on mean flows. Compared with its predecessor, the flow matching-based model, MeanSE trains the neural network to model the average velocity field rather than the instantaneous velocity field, thus providing better performance on 1-NFE condition. Experimental results demonstrate that our proposed MeanSE model not only shows stronger 1-NFE enhancement performance, but also comprehensively surpasses the flow matching-based model in generalization capabilities. In future work, we plan to apply mean flows to more speech processing tasks.

\bibliographystyle{IEEEbib}
\ninept
\clearpage
\bibliography{refs}

\end{document}